\documentclass[%
 reprint,
 amsmath,amssymb,
 aps,
]{revtex4-2}
\usepackage{color}
\usepackage{graphicx}
\usepackage{dcolumn}
\usepackage{bm}
\usepackage[mathlines]{lineno}
\linenumbers\relax 
\usepackage{amsfonts}
\usepackage{sidecap}
\usepackage{float}
\usepackage{parskip}
\usepackage{grffile}
\usepackage{comment}

\usepackage{color}
\usepackage{hyperref}
\usepackage{hhline}
\usepackage{mathtools}
\usepackage[normalem]{ulem}
\makeatletter
\def\@eqnnum{{\normalsize \normalcolor (\theequation)}}
 \makeatother
\begin{document}\nolinenumbers
\title{Onset of synchronization in contrarians with higher-order interactions}
\author{Vasundhara Rathore$^{1}$}
\author{Ayushi Suman$^{2}$}
\author{Sarika Jalan$^{2}$}\email{sarikajalan9@gmail.com}
\affiliation{1. Department of Biosciences and Biomedical Engineering, Indian Institute of Technology Indore, Khandwa Road, Simrol, Indore-453552, India}
\affiliation{2. Department of Physics, Indian Institute of Technology Indore, Khandwa Road, Simrol, Indore-453552, India}

\date{\today}

\begin{abstract}
We investigate the impact of contrarians (via negative coupling) in a multilayer network of phase oscillators having higher-order interactions. We show that the multilayer framework facilitates synchronization onset in the negative pairwise coupling regime.
The multilayering strength governs the onset of synchronization and the nature of the phase transition, whereas the backward  critical couplings depend on higher-order interaction strength. The system does not synchronize below a critical value of multilayering strength. 
The numerical results agree with the analytical predictions using the Ott-Antonsen approach. The results presented here may be useful for understanding emergent behaviors in real-world complex systems with contrarians and higher-order interactions, such as the brain and society.
\end{abstract}

\pacs{05.45.-a,89.75.-k,05.45.Xt}
\keywords{}
\maketitle

\paragraph{\bf{Introduction:}}
Many real-world systems inherently possess higher-order interactions beyond traditional pair-wise interaction setups.
Examples incorporate collaboration graphs \cite{collab_eg}, coauthorship data, music collaboration data (nodes are rap artists, simplices are sets of rappers collaborating on songs) \cite{pnas_HO}, cliques and cavities in the human connectome \cite{brain_HO}, etc.
Modeling dynamical systems incorporating higher-order interactions has revealed many emerging phenomena that may not be evident if only pair-wise interactions  are considered \cite{AD_1,AD_2,skardal_struct}. In 2011, Tanaka and Aoyagi investigated the ensemble dynamics of phase oscillators with three-body interactions {\cite{tanaka}}. They reported multistability for neuronal networks with three-body interaction, which means that as the initial condition is varied, the number of synchronized neurons at the steady state varies. Later, studies on globally coupled oscillators on simplicial complexes, in the absence of pair-wise interactions, \cite{abrupt_desync} revealed an abrupt desynchronization  without its counterpart of abrupt synchronization transition. Further investigations of systems involving both pairwise and higher-order interaction \cite{nature_skardal}, \cite{dib_ghosh1,dib_ghosh2}, real-world networks with higher-order interaction \cite{social_networks}, attractive-repulsive coupling have unveiled a plethora of interesting behaviors \cite{review_article}.
Furthermore, for pairwise coupling case, it is known that the oscillators negatively coupled to the mean field (contrarians) tend to align in anti-phase with the mean-field and suppress the global synchronization \cite{strogatz_paper,mp_ES_frust, chimera_rep, inhibition_1, inhibition_2}. Positive-negative interactions in a system are analogous to the model of spin glasses \cite{spinglass}. Frustrated interactions exist in a spin glass because of the random alignment of the spin. Ferromagnetic bonds exist (neighbors have the same orientation) and antiferromagnetic bonds (neighbors have exactly the opposite orientation). Similar to the ferromagnetic interaction, positive coupling favors the alignment of the oscillators in phase. In contrast, similar to anti-ferromagnetic interactions, negative couplings attempt to set oscillators apart and favor a phase difference of $\pi$. 
People have analyzed both real and synthetic ecological networks where facilitation and competition interactions co-exist \cite{ecological_example, ecological_example_1, ecological_example_2, ecological_example_3}. Other examples of positive and negative interactions are the social model of opinion formation dynamics \cite{poll_example}, inhibitory-excitatory neurons in the brain \cite{brain}, and epithelial-mesenchymal cells in the mouse skin \cite{haircycle}.
It has been recently shown that the contrarians can  synchronize beyond the pairwise interactions in the presence of higher-order interactions \cite{nature_skardal,bocc_pos_neg}. The higher-order interactions can instigate the emergence of a coherent state in a single-layer network even when the oscillators are coupled negatively to the mean field by  stabilizing the synchronized state.
Moving to the multilayer framework is crucial in characterizing the interactions among the component of various complex systems which single layer framework cannot capture \cite{multi_def, mp_tune, mp_so, strelkova, makov, anil, aradhana, aradhana2, prio}. Multilayer networks embody different  connectivity channels, and a layer represents each channel \cite{multi_appl}. For example,  in transportation networks, different layers represent different modes of transport, like bus, train, air, etc., between different cities \cite{transport}.  Studies on coupled phase oscillators on simplicial complexes on  multilayer systems  have revealed multiple basins of attraction and routes to abrupt first-order transition to synchronization \cite{ayushi_paper}.
Further, when a layer of phase oscillators with positive pairwise connections multilayered with another layer with negative pairwise connections, such a multilayering leads to explosive synchronization transition in both the layers \cite{ES_vasu_PRE} with the onset of the transition lying at the positive value. The current work studies multilayer networks of  hypergraphs with negative couplings. 
Here we investigate how contrarian in one layer affects the dynamical evolution of contrarians and antagonists in another layer but in the presence of higher-order interactions. To model this, we fix one layer represented by the Kuramoto model with pairwise (i.e. 1-simplex) negative interactions, and another layer is set up to have triadic (i.e. 2-simplex) positive and negative interactions, and the multilayering can also take a positive or negative value. There could be a scenario when all the interactions in the system are negative. We report the onset of synchronization at negative coupling strength. The multiplexing strength governs the nature of the transition from an incoherent state to a completely synchronized state and the critical coupling strength at which the transition occurs. 
First, we show that for fixed higher-order coupling i.e. $K_{t}^{(1)}>3$, the first-order transition is obtained and the onset of synchronization is obtained at negative $K_{pcf}^{(2)}$ depending on the magnitude of $D_{x}$(analytically and numerically). Then, we show how hysteresis width depends on the magnitude of $D_{x}$ and even for strong negative $K_{p}^{(2)}$ synchronization persists in the system as we increase the magnitude of $D_{x}$. Next, we elaborate on the role of higher-order coupling strength $K_{t}^{(1)}$. The nature of phase transition from second-order to first-order is governed by $K_{t}^{(1)}$ as well as backward critical coupling strength $K_{pcb}^{(2)}$ depends on $K_{t}^{(1)}$. Further, we see the behavior of $r^{(1)},r^{(2)}$ with the change in $K_{t}^{(1)}$ and for fix pair-wise negative coupling $K_{p}^{(2)}=-2$. A bifurcation is obtained at $D_{x}=2.8$ where $r^{(1),(2)}=0$ no more remains the stable solution, and a second-order transition to synchronization is obtained, and further with the increase in $D_{x}$, the oscillators remain in the synchronized state even for negative $K_{t}^{(1)}$.   

\paragraph{\bf{Model:}}
We consider an undirected and unweighted bilayer network system with each layer consisting of $N$ Kuramoto oscillators. The coupled dynamics is governed by:
\begin{equation}\label{eqn1}
\dot\theta_{i}^{(l)}=\omega_{i}^{(l)}+ H_{i}^{(l)} + \frac{D_x}{N} \sum_{j=1}^N\sin(\theta_{j}^{(l^{'})}-\theta_{i}^{(l)}),
\end{equation}
where,
\begin{eqnarray}
    H_{i}^{(l)}&=&\frac{K_{t}^{(l)}}{N^2} \sum_{j=1}^N \sum_{k=1}^N \sin(2\theta_{j}^{(l)}-\theta_{k}^{(l)}-\theta_{i}^{(l)})\nonumber\\ &+&  \frac{K_{p}^{(l)}}{N} \sum_{j=1}^N\sin(\theta_{j}^{(l)}-\theta_{i}^{(l)}) \nonumber
\end{eqnarray}
Here, $l$ and $l^{'}$ represent the layer index, when $l=1$ $l^{'}=2$ and vice-versa. For layer $1$ ($l=1$), $K_{p}^{(1)}=0$ and for layer $2$ ($l=2$), $K_{t}^{(2)}=0$. $K_{t}^{(1)}$ is the coupling between 2-simplex (triadic) interaction, and $K_{p}^{(2)}$ is the coupling between pairwise interaction. $\omega_{i}^{(1)}$, $\omega_{i}^{(2)}$ are the intrinsic frequencies of layer $1$ and $2$ respectively. $D_{x}$ is the multiplexing strength. 

The measure of the degree of synchronization, a generalized complex order parameter, is defined
\begin{eqnarray}
        r^{(l)}_q e^{\iota\Psi^{(l)}_q} = \Bigg|\frac{1}{N}\sum_{j=1}^N e^{\iota q\theta_{j}^{(l)}}\Bigg|\quad
\end{eqnarray}
This quantity represents a vector sum of all the phase oscillators on the unit circle. The superscript in parentheses denotes the layer index(unless stated otherwise), and the subscript $'q'$ represents the moment of the quantity defined. The first moment measures one-cluster synchronization; the second moment measures the extent of two-cluster synchronization, and so on. Evidently, $r^{(l)}=1$ implies that all the oscillators are in a coherent state, whereas $r^{(l)}=0$ represents that all the oscillators are in an incoherent state.  

\begin{figure}
	\centering
	\includegraphics[width=1.0\columnwidth,height=5.3 cm]{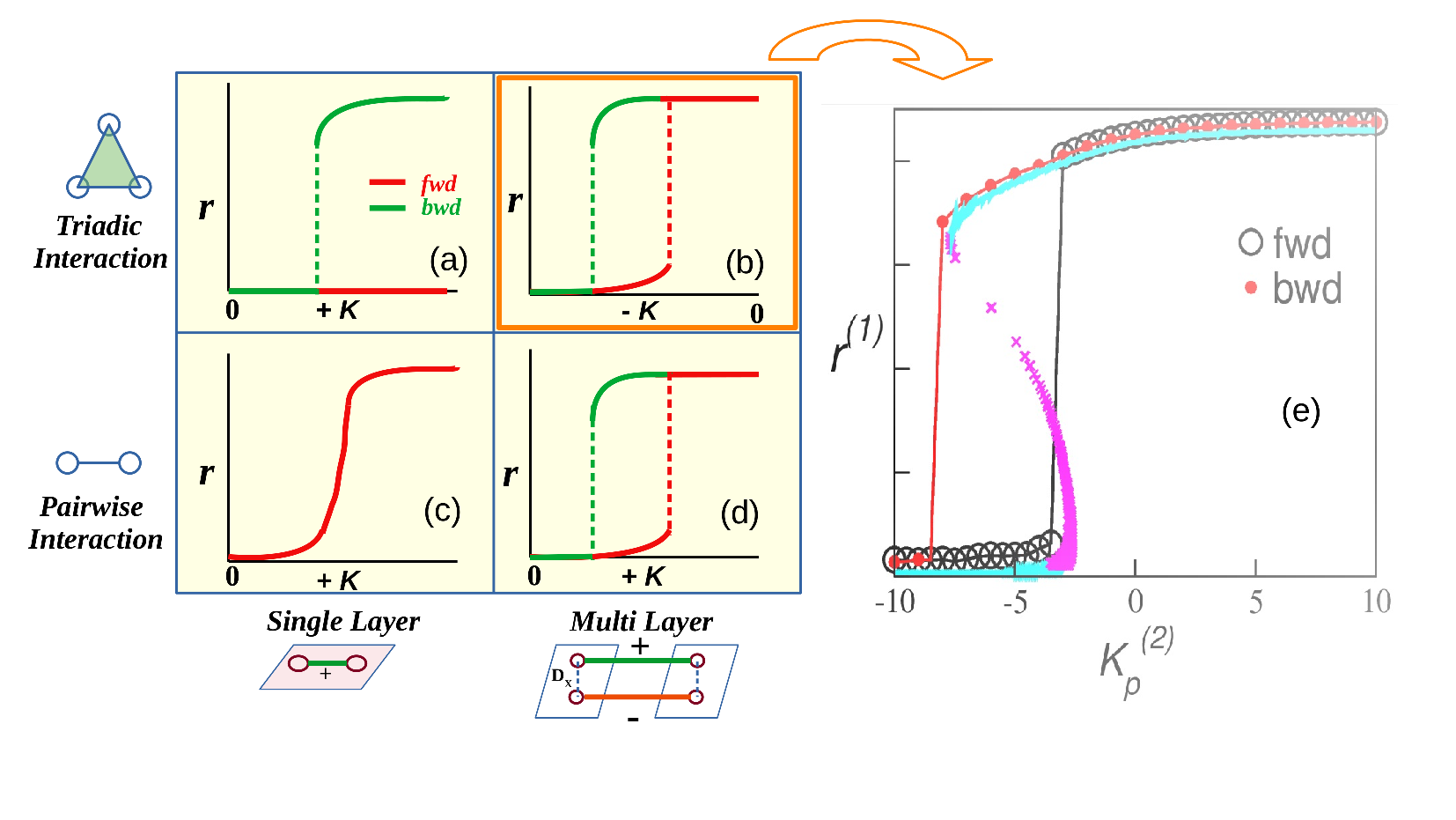}\\
	\caption{(Color Online) Schematic diagram for order parameter behavior as a function of coupling strength. (a) A single-layer network with triadic interaction showing only an abrupt jump in backward transition without any counter forward jumps, (b) the hysteresis shifts to the negative regime when one layer with triadic interaction is multiplexed with another layer having only pairwise interaction, (c) single layer with pairwise interaction depicts the continuous transition to synchronization, (d) when one layer with positive pairwise interaction is multiplexed with another layer with negative pairwise interaction, hysteresis is obtained in the positive regime of pairwise coupling. (e) Numerical and analytical representation of $r^{(1)}$ as a function of $K_{p}^{(2)}$ for $D_{x}=3$ and $K_{t}^{(1)}=6$, $N=3000$.}
	\label{Schematic}
\end{figure}

\paragraph{\bf{Analytical Calculations:}}
Writing Eq.~\ref{eqn1} into the mean-field form,
\begin{align}\label{eqn3}
\begin{split}
\dot\theta_{i}^{(1)} = \omega_{i}^{(1)}+ &K_{t}^{(1)} r_{1}^{(1)} r_{2}^{(1)} \sin(\Psi_{2}^{(1)}-\Psi_{1}^{(1)}-\theta_{i}^{(1)})\\+ & D_x r_{1}^{(2)} \sin(\Psi_{1}^{(2)}-\theta_{i}^{(1)}),
\\ \dot\theta_{i}^{(2)} = \omega_{i}^{(2)}+ &K_{p}^{(2)} r_{1}^{(2)} \sin(\Psi_{1}^{(2)}-\theta_{i}^{(2)})\\+ &D_x r_{1}^{(1)} \sin(\Psi_{1}^{(1)}-\theta_{i}^{(2)})
\end{split}
\end{align}
where $(r_{q}^{(1)}, \Psi_{q}^{(1)})$, $(r_{q}^{(2)},\Psi_{q}^{(2)})$ represent order parameter and mean phase of layer $1$ and $2$ respectively with $q=1,2$.
The density function $f_{(l)}(\theta,\omega,t)$ for the $l^{th}$ layer obeys the continuity equation,
\begin{equation}\label{eqn7}
\frac {\delta f_{(l)}}{\delta t}+ \frac {\delta(f_{(l)}\dot\theta^{(l)})}{\delta \theta^{(l)}}=0
\end{equation}
Expanding density function into Fourier series,
\begin{equation}
f_{(l)}(\theta,\omega,t)=\frac{g(\omega)}{2\pi} (1+\sum_{n=1}^\infty f_{n}(\omega,t) e^{in\theta}+c.c.)
\end{equation}
where c.c. is the complex conjugate of the first term. Using Ott-Antonsen ansatz, $f_{n}(\omega,t)=[\alpha(\omega,t)]^{n}$, where $|\alpha(\omega,t)\le1|$.
Substituting this series expansion into \eqref{eqn7}and following the \cite{ayushi_paper}, the system of $2N$ equation reduces to a system of 4 coupled differential equations (From now onwards, $r^{(1),(2)}$ represent the global order parameter which calculates the one-cluster state of layer $1$ and $2$ respectively).
\begin{align}\label{eqn11}
\begin{split}
\Dot{r}^{(1)}=-\Delta r^{(1)} + &\frac{K_{t}^{(1)}}{2} [(r^{(1)})^{3}-(r^{(1)})^{5}] -\frac {D_{x}}{2}[r^{(2)} (r^{(1)})^{2}\\-r^{(2)}] \\
\Dot{\Psi}^{(1)}=\omega_0^{(1)} + &\frac{D_{x}}{2} [\frac{r^{(2)}}{r^{(1)}}+r^{(2)} r^{(1)}] \sin(\Psi^{(1)}-\Psi^{(2)})
\\
\Dot{r}^{(2)}=-\Delta r^{(2)} + &\frac{K_{p}^{(2)}}{2} [r^{(2)}-(r^{(2)})^{3}] -\frac {D_{x}}{2}[r^{(1)} (r^{(2)})^{2}\\-r^{(1)}]
\\
\Dot{\Psi}^{(2)}=\omega_0^{(2)} + &\frac{D_{x}}{2} [\frac{r^{(1)}}{r^{(2)}}+ r^{(1)} r^{(2)}] \sin(\Psi^{(2)}-\Psi^{(1)})
\end{split}
\end{align}

\begin{figure}
\centering
\includegraphics[width=1.0\columnwidth, height=7cm]{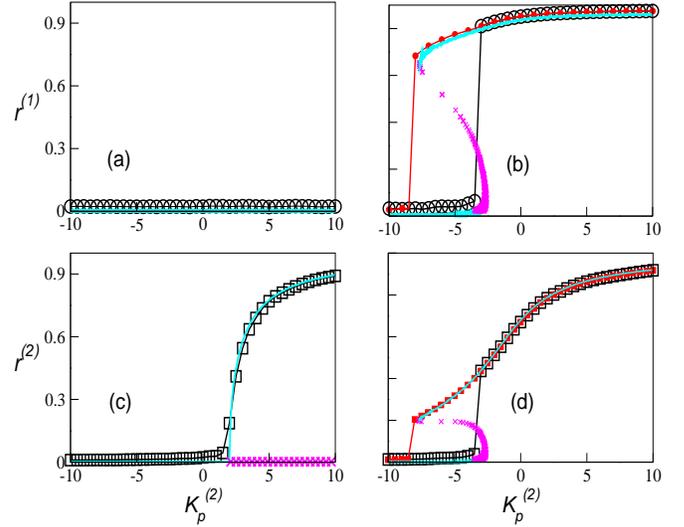}\\
	\caption{(Color Online) Shifting of the whole hysteresis in the negative regime of $K_{p}^{(2)}$ with change in multiplexing strength $D_{x}$ for fix $K_{t}^{(1)}=6$. The figure depicts the behavior of order parameter of layer $1$ and $2$ (i.e. $r^{(1)}$, $r^{(2)}$) with $K_{p}^{(2)}$. Cyan solid line (stable) and magenta crossed the line (unstable) represent the analytical curve obtained from equation \eqref{eqn11}, whereas black and red circles represent the forward and backward transitions obtained numerically. The network size is $N=3000$. (a) and (c) represent the forward transition of layer $1$ and $2$ respectively, with change in $K_{p}^{(2)}$ for $D_{x}=0$, (b) and (d) represent forward and backward transition of layer $1$ and $2$ respectively with change in $K_{p}^{(2)}$ for $D_{x}=3$.}
	\label{hysteresis_shifting}
\end{figure}

For any steady state, it is required that $\Dot{r}^{(1)}=\Dot{r}^{(2)}=\Dot{\Psi}^{(1)}=\Dot{\Psi}^{(2)}=0$. For $\Dot{\Psi}^{(1)},\Dot{\Psi}^{(2)} $ to be zero, $\Psi^{(1)}-\Psi^{(2)}$ must be $0$ or $\pi$ indicating the in-phase and out of phase synchronization between the two layers. In either case, the equation for $r^{(1)}$ and $r^{(2)}$ uncouple from the equation for $\Psi^{(1)}$ and $\Psi^{(2)}$. As a result, a steady-state stability analysis requires only two of the above differential equations.
\begin{align}
\begin{split}
g_1=\Dot{r}^{(1)}=-\Delta r^{(1)} &+ \frac{K_{t}^{(1)}}{2} [(r^{(1)})^{3}-(r^{(1)})^{5}] - \frac {D_{x}}{2} \\ [r^{(2)} (r^{(1)})^{2}-r^{(2)}] = 0 \\
g_2=\Dot{r}^{(2)}=-\Delta r^{(2)} &+ \frac{K_{p}^{(2)}}{2} [r^{(2)}-(r^{(2)})^{3}] -  \frac {D_{x}}{2} \\ [r^{(1)} (r^{(2)})^{2}-r^{(1)}] = 0
\end{split}
\end{align}
it can be seen that $(r^{(1)^*},r^{(2)^*})=(0,0)$ is always a steady state for any parameter value. The other steady state, a function of all the parameters, may not be found analytically. Still, a stability analysis of $(r^{(1)^*},r^{(2)^*})=(0,0)$ state will provide a constraint on the parameters and will yield the regime for which synchronization can be observed for negative coupling. The characteristic equation 
\begin{align}
\begin{split}
    |J-\lambda X| &= 0 \\
    with \;   J &=
{\begin{vmatrix}
\frac{\delta g_1}{\delta r^{(1)}} & \frac{\delta g_1}{\delta r^{(2)}} \\[6 pt]
\frac{\delta g_2}{\delta r^{(1)}} & \frac{\delta g_2}{\delta r^{(2)}}
\end{vmatrix}}_{r^{(1)^*}=0,r^{(2)^*}=0}\\
&= 
\begin{vmatrix}
-\Delta & \frac{D_{x}}{2} \\[6 pt]
\frac{D_{x}}{2} & -\Delta + \frac{K_{p}^{(2)}}{2}
\end{vmatrix} 
\end{split}
\end{align}
will thus give the eigenvalues
\begin{equation*}
    \lambda= \frac{1}{2}\left[-\beta\pm \sqrt{\beta^2-4\left(\Delta^2-\Delta\frac{K_{p}^{(2)}}{2}-\frac{D_{x}^{2}}{4}\right)}\right]
\end{equation*}
with $\beta=\left(2\Delta-\frac{K_{p}^{(2)}}{2}\right)$. If both the eigenvalues are negative, the incoherent state is stable, but if one of the eigenvalues turns positive for some parameter values, the state will become saddle. That parameter value will correspond to the onset of synchronization. Henceforth,
\begin{equation*}
    -\beta > \sqrt{\beta^2-4\left(\Delta^2-\Delta\frac{K_{p}^{(2)}}{2}-\frac{D_{x}^{2}}{4}\right)}
\end{equation*}

\begin{figure}
	\centering
\includegraphics[width=1.0\columnwidth]{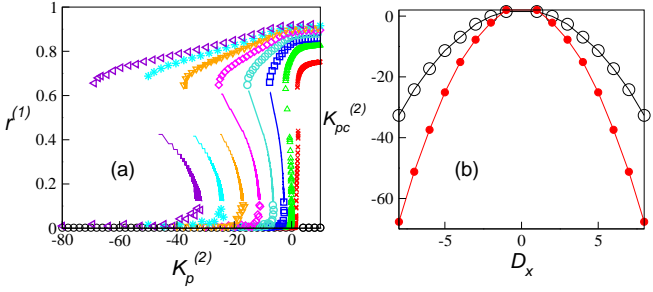}\\
	\caption{(Color Online) Forward and backwards critical coupling strengths as a function of $D_{x}$ and pair-wise coupling strength for fixed $K_{t}^{(1)}=6$. (a) Analytical presentation of $r^{(1)}$ as a function of $K_{p}^{(2)}$ for $D_{x}$ ranging from 0 to 8 (black circles to violet left triangles). (b) Forward critical coupling strength $K_{pcf}^{(2)}$ (black empty circle) and backward critical coupling strength $K_{pcb}^{(2)}$ (red filled circle) as a function of $D_{x}$.}
	\label{Dx_effect}
 \end{figure}

which gives a bound on $K_{p}^{(2)}$ for the system to leave the incoherent state through a subcritical pitchfork bifurcation. The above equation simplifies to
\begin{equation}\label{eqn12}
    K_{p}^{(2)}>\frac{4\Delta^2-D_{x}^{2}}{2\Delta}
\end{equation}

Thus, the onset of synchronization depends on the spread of natural frequencies, which is natural as for a larger spread in natural frequencies, a lesser fraction of oscillators participates in the synchronized cluster. Interestingly, the onset of synchronization also depends on only the magnitude of $D_x$. For $D_x=0$, synchronization occurs at $2\Delta$, and by increasing $D_x$, the critical point (transition point) shifts towards the left. For $\pm D_{x}>2\Delta$ $K_{pc}^{(2)}$ becomes negative, making the onset of synchronization for repulsive coupling strength. Note that the critical point is independent of the higher-order coupling strength of the other layer. Ergo, synchronization can manifest for multilayer systems with only inhibitory couplings at all levels. Next section presents simulation results along with the semi-analytical prediction \ref{eqn11} for the whole $r^{(1)},r^{(2)}-K_{t}^{(1)}, K_{p}^{(2)}$ space, respectively.

We consider a bi-layer network with one-to-all multiplexing, with a globally coupled system representing both layers. To investigate the synchronization profile of each layer, we calculate the order parameters $r^{(1)}$ and $r^{(2)}$ as a function of intra-layer coupling strengths. To introduce contrarians in either of the layers $K_{t}^{(1)}$ or $K_{p}^{(2)}$ is set to be negative accordingly. Starting from the random initial condition $K_{t}^{(1)}$ (or $K_{p}^{(2)}$) is increased adiabatically from an incoherent to a coherent state, which is called a forward transition. Then, the coupling strength $K_{t}^{(1)}$ (or $K_{p}^{(2)}$) is decreased adiabatically from the obtained coherent state to the incoherent state to analyze the backward transition. In case there is no forward transition, to realize the backward transition, we start from the initial condition with all the oscillators lying in the coherent state (i.e., all oscillators have either phase $0$ or $\pi$) and then adiabatically reduce the coupling strength with a step size of $\delta K_{t}^{(1)}$ (or $\delta K_{p}^{(2)}$) until the incoherent state. The initial frequency distribution is chosen to be Lorentzian with $\sigma=1$. The phase evolution of \ref{eqn1} is integrated using the Runge–Kutta 4th order method with step size $dt = 0.01$ for a long enough time to arrive at a stationary state after discarding the initial transients. 

 \begin{figure}
	\centering
\includegraphics[width=1.0\columnwidth]{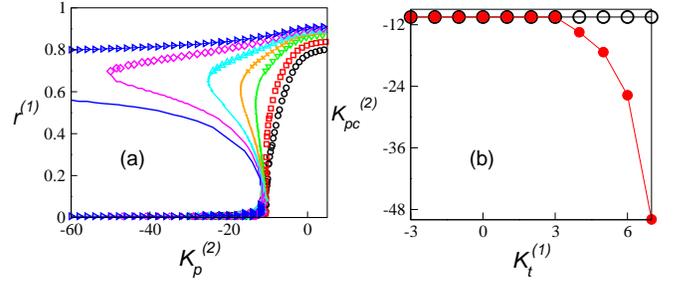}\\
	\caption{(Color Online) $K_{pcb}^{(2)}$ as a function of pair-wise and triadic coupling strengths, respectively,  by fixing $K_{pcf}^{(2)}=-10.5$. (a) Analytical presentation of $r^{(1)}$ as a function of $K_{p}^{(2)}$ for $D_{x}=5$ and $K_{t}^{(1)}=0$ (black circle), $2$ (red square), $4$ (green triangle down), $5$ (orange cross), $6$ (cyan triangle up), $7$ (magenta diamond) and $8$ (blue triangle right). (b) Forward critical coupling strength $K_{pcf}^{(2)}$ (black empty circle) and backward critical coupling strength $K_{pcb}^{(2)}$ (red filled circle) as a function of $K_{t}^{(1)}$ for $D_{x}=5$.}
	\label{k1_effect}
\end{figure}
\paragraph{\bf{Onset of synchronization at negative $K_{p}^{(2)}$:}} Fig.~\ref{hysteresis_shifting} illustrates the shifting of the whole hysteresis in the negative regime of $K_{p}^{(2)}$ as $D_{x}$ is varied.  For $D_{x}=0$ ($(a)$ and $(c)$), 
layer $1$ does not manifest synchronization as $K_p^{(2)}$ is increased and layer $2$ experiences the second-order transition to synchronization. However, as soon as $D_{x}$ is increased to $3$, we witness the onset of synchronization at negative $K_{pc}^{(2)}$ ($~-2.5$ from \eqref{eqn12}). For $D_{x}=3$, starting from the random initial condition, there exists a first-order phase transition with hysteresis, and the whole hysteresis is obtained in the negative regime of $K_{p}^{(2)}$ (Fig.~\ref{hysteresis_shifting}(b)). Layer $2$ also manifests a jump in $r^{(2)}$ from an incoherent to a partially synchronized state (with hysteresis). Particularly, the critical coupling strength at which forward transition is obtained ($K_{pcf}^{(2)}$) depends on $D_{x}$. Still, the nature of transition, if it will be first-order or second-order, depends on the value of $K_{t}^{(1)}$. For $(D_{x}=3)$, $K_{t}^{(1)}>2$ first-order transition to synchronization is obtained, whereas for $K_{t}^{(1)}<2$ second-order transition is obtained. Further, for $D_{x}>3$ even if $K_{t}^{(1)}$ has a high negative value, the onset of synchronization is obtained in the negative regime of $K_{p}^{(2)}$. Note that even if both the layers are comprised of contrarians, independent of the sign of $D_{x}$ (positive or negative), for $|{D_{x}}|>2$, both the layers attain synchronization. 
Fig.~\ref{hysteresis_shifting} shows a good agreement between the numerical and analytical results.

\paragraph{\bf{Impact of multiplexing strength:}} $D_{x}$ plays a significant role in governing the dynamics of the entire system. Two key roles 
 played by $D_{x}$ are; (a) it  facilitates the onset of synchronization in the negative regime of $K_{p}^{(2)}$, (b) the width of hysteresis increases with $D_{x}$. Fig.~\ref{Dx_effect}(a) represents an analytical curve depicting the behavior of $r^{(1)}$ with respect to $K_{p}^{(2)}$ for different $D_{x}$ values at a fixed $K_{t}^{(1)}$. It can be seen that both $K_{pcf}^{(2)}$ and $K_{pcb}^{(2)}$  shift towards left with the increase in $D_{x}$ from $0$ to $8$. 
 For $D_{x}=8$ the onset of synchronization is obtained at a strong negative value of $K_{p}^{(2)}$ $(~-35)$ and a prolonged hysteresis is obtained in the range $-70<K_{pcf}^{(2)}<-35$. These results affirm that $D_{x}$ facilitates the synchronized state to exist for a larger range of negative $K_{p}^{(2)}$. It is clear from Fig.~\ref{Dx_effect} that both $K_{pcf}^{(2)}$ and $K_{pcb}^{(2)}$ shift with $D_{x}$ and hysteresis width increases with $D_{x}$. This shifting in both forward and backward critical coupling strength is symmetric around $D_{x}=0$, as the dynamics only depend on the magnitude of $D_{x}$ (not on its sign).
\begin{figure} 
	\centering
	\includegraphics[width=0.85\columnwidth]{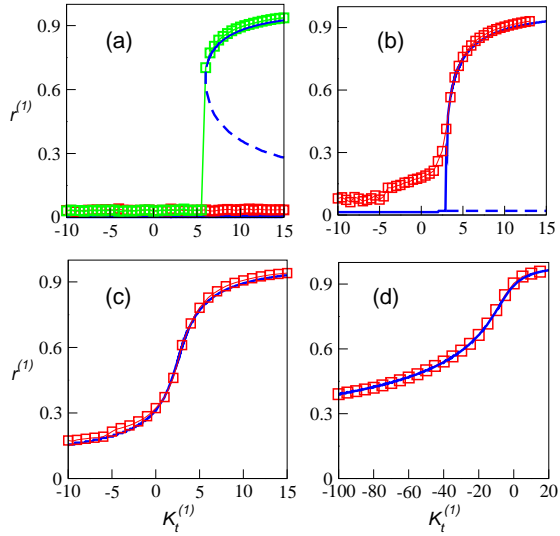}\\
	\caption{(Color Online) Change in the nature of phase transition of layer $1$ with $D_{x}$ for $K_{p}^{(2)}=-2$. $D_{x}=2.8$ is the bifurcation point. The blue line shows the analytical curve, square (red) represents the forward transition, square (green) represents the backward transition for (a) $D_{x}=2$, (b) $D_{x}=2.8$, (c) $D_{x}=3.0$ and (d) $D_{x}=10$.}
	\label{layer1_k1_change}
\end{figure}
\paragraph{\bf{Impact of higher-order coupling $K_{t}^{(1)}$:}} For any given $|{D_{x}}|>2$, $K_{t}^{(1)}$ governs the nature of transition and associated bifurcation. 
Fig.~\ref{Dx_effect}(a) depicts 
 that $K_{pcf}^{(2)}$ does not depend upon $K_{t}^{(1)}$, however, $K_{pcb}^{(2)}$ depends on $K_{t}^{(1)}$ as it keeps shifting towards left with an increase in $K_{t}^{(1)}$ thereby yielding a prolonged hysteresis. For $K_{t}^{(1)}=8$, the width of hysteresis increases remarkably. Hence the higher-order coupling supports the synchronization to persist even for a long range of negative $K_{p}^{(2)}$. Fig.~\ref{Dx_effect}(b) 
 depicts that $K_{pcf}^{(2)}$ remains fixed to  -10.5 for all $K_{t}^{(1)}$ values  (positive or negative). For $K_{t}^{(1)}>3$, the nature of transition changes from the second-order to the first-order. Notably,  even for negative $K_{t}^{(1)}$ values, the system attains synchronization via the second-order route.
\begin{figure}
	\centering
	\includegraphics[width=0.95\columnwidth]{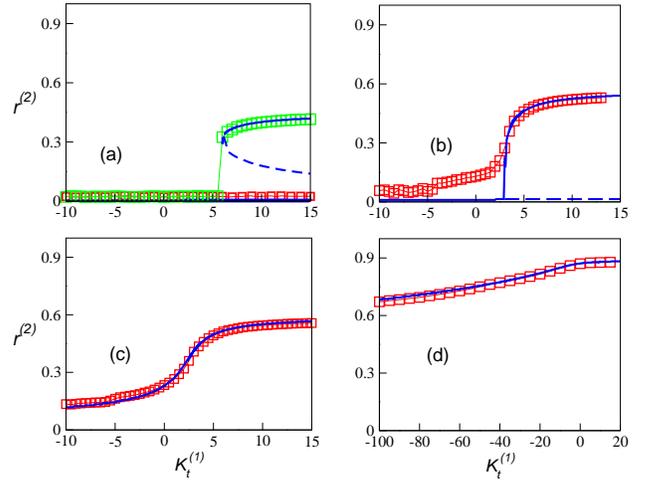}\\
	\caption{(Color Online) Change in the nature of phase transition of layer $2$ with $D_{x}$ for $K_{p}^{(2)}=-2$. $D_{x}=2.8$ is the bifurcation point.The blue line shows the analytical curve, square (red) represents the forward transition, square (green) represents the backward transition for (a) $D_{x}=2$, (b) $D_{x}=2.8$, (c) $D_{x}=3.0$ and (d) $D_{x}=10$.}
	\label{layer2_k1_change}
\end{figure}
\paragraph{\bf{Persistence of synchronization against $K_{t}^{(1)}$:}} Next, by fixing $K_{p}^{(2)}=-2$ we analyze $r^{(1)}$ by changing  $K_{t}^{(1)}$, and  how its profiles changes for different $D_{x}$ values. For $D_{x}=2$ and $K_{p}^{(2)}=-2$, starting from the random initial conditions, there is no forward transition to synchronization, i.e. $r=0$ remains the stable solution for all $k_t^{(1)}$. Upon setting the initial condition for oscillators having equal phases (i.e. $\theta=0$ or $\pi$), during the backward transition, an abrupt jump is obtained from a coherent to an incoherent state (Fig.~\ref{layer1_k1_change} (a)). As $D_{x}$ is increased further from $2$, the unstable branch keeps approaching the $r=0$ stable branch, and at $D_{x}~2.8$ the unstable branch merges with the stable $r=0$ branch yielding a continuous transition to synchronization. $D_{x}=2.8$ is the bifurcation point. For a further increase in $D_{x}$ (Fig.~\ref{layer1_k1_change} (c)), there is a continuous transition to synchronization; however, up to negative values of $K_{t}^{(1)}$ $(~-10)$, $r^{(1)}$ does not reach $0$.
Further for $|{D_{x}|}=10$,  $r^{(1)}$ indicates partial synchronization even for high negative values of $K_{t}^{(1)}$ (Fig.~\ref{layer1_k1_change} (d)). Here also, synchronisation persists in the system even when both layers consist of contrarians.
Next, moving forward to explain the behaviours of layer $2$, for $|{D_{x}|}=2$,  an abrupt jump to a partially synchronized state is obtained with no counter forward synchronization transition (Fig.~\ref{layer2_k1_change} (a)). At $D_{x}=2.8$, the bifurcation occurs, the unstable branch merges with the stable one, and the second-order transition to a partially synchronized state is obtained (Fig.~\ref{layer2_k1_change} (b) and (c)). The interesting point here is that for $|{D_{x}|}=10$, even for high negative $K_{t}^{(1)}$, $r^{(2)}$ keeps attaining a large  ($0.7$) (Fig.~\ref{layer2_k1_change} (d)) even stronger synchronization than that of layer $1$. Ergo, while the oscillators with pairwise negative coupling do not show synchronization, we witness  a strong synchronization facilitated by multilayer and higher-order interactions.

\paragraph{\bf{Conclusion:}}
We studied multilayer networks with positive and negative pair-wise and triadic interactions and investigated the impact of contrarians in one layer on the dynamical  evolution of agonist and contrarian in the other layer. 
We found that multilayering of the contrarian layer with another protagonist or contrarian layer having higher-order interactions facilitates the onset of first-order synchronization at negative coupling ($K_{p}^{(2)}<0$). We analytically calculated the bounds for $D_x$, which aids the transition to synchronization at negative coupling strength. 
With an increase in $|{D_{x}}|$, the forward as well as backward critical points ($K_{pc}^{(2)}$) keep shifting in the negative direction, thereby resulting in the increment of hysteresis width demonstrating aid of $D_{x}$ in synchronization of contrarians. Further, for a fixed $D_x$ value, the nature of the transition is decided by triadic coupling strength. The forward critical pairwise coupling depends only on $D_{x}$ and not on triadic coupling,  whereas the backward critical pairwise coupling is governed by triadic interactions. Therefore, one can tune $D_{x}$ and $K_{t}^{(1)}$ such that synchronization persists for a larger range of negative $K_{p}^{(2)}$. Furthermore, if the pairwise coupling strength is fixed to a negative value, there exists a threshold value of $D_{x}$ at which bifurcation occurs and $r=0$ does not remain a stable solution any longer, and, a second-order transition to synchronization is obtained (even when both the layers comprised of contrarians). Both the analytical and numerical results match well. Our study provides a method to increase the persistence of synchronization in contrarians in multilayer systems having higher-order interactions. A straightforward extension of the work is to include network architecture and investigate how structural properties of hypergraphs affect the whole dynamics.

\section{Acknowledgment} SJ gratefully acknowledges SERB Power grant SPF/2021/000136. The work is supported by the computational facility received from the Department of Science and Technology (DST), Government of India under FIST scheme (Grant No. SR/FST/PSI-225/2016). VR is thankful to Govt of India, DST grant DST/INSPIRE Fellowship/[IF180308].

\end{document}